\documentclass[twocolumn,aps,pre,floatfix,superscriptaddress,longbibliography]{revtex4-1}
\pdfoutput=1 
\usepackage{amsmath,amssymb,eucal,graphicx,float,epstopdf,cases,epigraph}
\usepackage{epsfig}
\usepackage[utf8]{inputenc}
\usepackage{textcomp,csquotes}

\usepackage[colorlinks=true, urlcolor=blue, anchorcolor=blue, citecolor=blue,filecolor=blue,linkcolor=blue,menucolor=blue]{hyperref}

\newcommand{\ii}{{\rm i}}

\newcommand{\ra}{\rangle}
\newcommand{\la}{\langle}
\newcommand{\bea}{\begin{eqQnarray}}
	\newcommand{\eea}{\end{eqnarray}}
\newcommand{\beq}{\begin{equation}}
	\newcommand{\eeq}{\end{equation}}
\newcommand{\be}{\begin{equation}}
	\newcommand{\ee}{\end{equation}}
\newcommand{\beqa}{\begin{eqnarray}}
	\newcommand{\eeqa}{\end{eqnarray}}

\newcommand{\p}{\partial}
\newcommand{\Hil}{\mathcal{H}}
\newcommand{\Vil}{\mathcal{V}}

\begin{document}
\title{Dynamical fluctuations in the Riesz gas}

\author{Rahul Dandekar}
\affiliation{Institut de Physique Th\'{e}orique, CEA/Saclay, F-91191 Gif-sur-Yvette Cedex, France}
\author{P. L. Krapivsky}
\affiliation{Department of Physics, Boston University, Boston, Massachusetts 02215, USA}
\affiliation{Santa Fe Institute, Santa Fe, New Mexico 87501, USA}
\author{Kirone Mallick}
\affiliation{Institut de Physique Th\'{e}orique, CEA/Saclay, F-91191 Gif-sur-Yvette Cedex, France}

\begin{abstract}
We consider an infinite system of particles on a line performing identical Brownian motions and interacting through the $|x-y|^{-s}$ Riesz potential, causing the
over-damped motion of particles. We investigate fluctuations of the integrated current and the position of a tagged particle. We show that for $0 < s < 1$, the standard deviations of both quantities grow as $t^{\frac{s}{2(1+s)}}$. When $s>1$, the interactions are effectively short-ranged, and the universal sub-diffusive $t^\frac{1}{4}$ growth emerges with only amplitude depending on the exponent. We also show that the two-time correlations of the tagged-particle position have the same form as for fractional Brownian motion.
\end{abstract}

\maketitle

\section{Introduction}
\label{sec:Intro}

Systems of diffusive particles interacting via short-ranged interactions have been actively investigated in the past few decades. Among the most popular research subjects is the emergence of the hydrodynamic behavior and large deviations in such systems \cite{Anna,Spohn91,Kipnis99,Derrida07,MFT15}. The equilibrium properties of systems with long-ranged interactions have also been studied \cite{Ruffo09,Ruffo14,Cohn19}, and in a few cases, their dynamical properties have been explored \cite{Slava96,Ginzburg97,marcos,gonzalez}. 

In this work, we consider particles on an infinite line interacting via a long-ranged Riesz potential \cite{Riesz38}
\begin{equation}
\label{Riesz-V}
V_{s}(x,y) = g\,s^{-1} |x-y|^{-s}
\end{equation}
The $s^{-1}$ pre-factor in \eqref{Riesz-V} is convenient to have as the derivative of the potential  that  drives the particles. We assume that both the exponent $s$ and the coupling constant $g$ are positive: $s>0$ and $g>0$. The potential is thus repulsive, and for any $s>0$, it is sufficiently strong so that particles cannot collide---diffusion cannot overwhelm the repulsion. The order of the particles never changes, so we have a single-file system.

The motion caused by the Riesz potential \eqref{Riesz-V} is assumed to be over-damped. Particles also undergo independent Brownian motions. This system can be thought of as a gas at a finite temperature, with (identical) diffusion coefficients proportional to the temperature.  Riesz gases with particles undergoing deterministic motion have been studied for a long time. Applications to astrophysics \cite{Chandra-43} where particles are interpreted as stars or galaxies, as well as applications to plasma physics \cite{Petrache17} are natural. Riesz gases also appear in the context of crystallization and packing problems \cite{Cohn17,Cohn19,Petrache20}, Ginzburg-Landau vortices \cite{Serfaty12}, and random matrices \cite{Forrester,Mehta}. Specific Riesz gases, most frequently Coulomb gases, appear in concrete applications. However, Riesz gases with $s \le 1$ can be experimentally engineered in cold atom systems \cite{chalony, zhang2017}, and they are potentially interesting in a view of applications to quantum computers. The zero-temperature dynamics of the Riesz gas demonstrates interesting properties, such as signatures of chaos \cite{Huse21}. In  mathematics, Riesz gases are also subject to intense  studies (see \cite{Serfaty21,Lewin22} for recent reviews).

Some Riesz gases have received special  attention. The interaction is logarithmic when $s\to 0$, and such Riesz gases are known as log gases
\cite{Forrester}. In one dimension, a log gas of Brownian motions is a Dyson gas \cite{Dyson62} introduced in the context of random matrices with eigenvalues playing the role of particles. In two dimensions, the log gas describes a genuine two-dimensional Coulomb interaction \cite{Lewin22,Chafai} and in the over-damped case with particles additionally performing two-dimensional Brownian motions, the gas is called the Ginibre gas \cite{Ginibre65,Burda14,Burda15}. The Coulomb gas in $d$ dimensions has the exponent $s=d-2$. The Calogero gas with $s = 2$ is mostly studied in one dimension \cite{Calogero69,Calogero71,Alexios06}, albeit it makes sense in arbitrary dimension.  Dipole-dipole interactions are anisotropic, but for dipoles confined in a 1D channel with identical orientation caused by an external magnetic field perpendicular to the channel, the (three-dimensional) interaction potential is the Riesz potential with $s=3$. This system has been studied experimentally \cite{Dipoles-1d}. We also mention that re-writing the Riesz potential as $(2a/|x-y|)^s$ shows that the gas of hard spheres with radii $a$ emerges in the $s \rightarrow \infty$ limit. 

In one dimension, equilibrium properties of a Riesz gas in a confining potential have been studied via the Coulomb-gas approach \cite{Kundu19,Kundu22a,Kundu22b,Beenaker,Flack} originally developed for the Dyson gas \cite{dean06,potters}. The equilibrium behavior changes qualitatively when the exponent passes through the threshold value $s=1$ corresponding to particles confined to a line but interacting through the (three-dimensional) Coulomb potential \cite{Serfaty17a,Serfaty18a,Kundu19}. For $s>1$, the gas is effectively short-ranged; for $0<s<1$, the gas is long-ranged and the free energy functional is non-local. 

The goal of the present work is to investigate {\em dynamical} properties of one-dimensional stochastic Riesz gases. Among a few studies of the dynamics of Riesz gases, we mention \cite{Spohn87,Spohn87b,Blaizot,Instanton14,Huse21,Burda14}. Still, little is known about dynamics and equilibrium properties of systems with finite number of particles in confining (usually harmonic) potential remains the most popular research area. 

Another feature of our work is reliance on the macroscopic fluctuating theory (MFT) \cite{Bertini01,Bertini02,Jordan03}. The MFT is a powerful deterministic framework derived from fluctuating hydrodynamics in the vanishing-noise limit. The MFT is widely applied to diffusive lattice gases with a single scalar field \cite{Derrida07,MFT15}. Extensions of the MFT to several interacting stochastic fields and to stochastic field theories with higher derivatives are also actively explored \cite{Pablo19b,Vilenkin16}. We show that similarly to the equilibrium properties, an MFT suitable to one-dimensional stochastic Riesz gases undergoes a qualitative change when the exponent passes through the threshold value $s=1$. 

In one dimension, the MFT allows the investigation of the statistics of quantities like the total current across the origin \cite{DG09,Meerson12a} and, for single-file systems, the total displacement of the tracer \cite{Kirone14,Kirone15a}. For systems with short-ranged interactions, the variance of both these quantities grows as $t^{\frac{1}{2}}$ for long times \cite{Harris1965,Arratia83,Kirone14,Kirone15a}.  An amusing subtlety of diffusive systems often present in one dimension concerns the initial conditions, viz., their ever-lasting nature \cite{Barkai}. If initial conditions are deterministic (also known as quenched), fluctuations are often different from fluctuations in random (also known as annealed) initial conditions. This is particularly striking for large deviations that can be much more probable (albeit still highly rare) in the annealed case. In higher than one dimension, fluctuations in deterministic and annealed settings are often identical in the leading order \cite{Meerson12b,Meerson14}. 

We now state the main results of this paper.  Starting with stochastic hydrodynamics of the one-dimensional Riesz gas we develop a deterministic reformulation analogous to the MFT of stochastic diffusive lattice gases. When $0<s<1$, the governing equations contain non-local terms, which did not appear in the original  MFT equations; when $s>1$, we recover the usual MFT equations with a density-dependent diffusion coefficient which we derive. Using the relevant MFT, we then  probe the asymptotic behavior of the variance of the integrated current $Q$ and of the position $X$ of the tracer. These asymptotic behaviors are obtained by performing a perturbation expansion \cite{Meerson12a} around the noiseless hydrodynamic solution that is just the steady state with uniform density $\rho$. As an expansion parameter, it is convenient \cite{Meerson12a} to use a Lagrange multiplier corresponding to $Q$  or $X$. The averages vanish, $\la Q\ra=\la X \ra=0$, in the uniform setting. The variances grow algebraically with time
\begin{equation}
\label{QX-gamma}
\la Q^2 \ra = \rho^2 \la X^2 \ra \sim t^{2 \gamma}
\end{equation}
with exponent depending on the Riesz exponent:
\begin{equation}
\label{gamma}
\gamma = 
\begin{cases}
	\frac{1}{2}\frac{s}{s+1}     & 0<s<1\\
	\frac{1}{4}                        &  s>1
\end{cases}
\end{equation}
In the marginal case of $s=1$ (which is physically important as it corresponds to particles confined to a line interacting through three-dimensional Coulomb potential), we argue that  $\la Q^2 \ra \sim \la X^2 \ra \sim \sqrt{t/\ln t}$. Note that  $Q$ and $X$ grow sub-diffusively with an  $s-$dependent exponent when $0<s<1$ and with the universal exponent $\frac{1}{4}$ when $s>1$. This value  $\frac{1}{4}$ is the same as that for short-ranged diffusive systems with forbidden overtaking such as simple exclusion processes \cite{Arratia83,KRB}.

We also determine the two-time correlation function of the position of the tracer (i.e., the tagged particle). The form of this function depends on the setting:
\begin{subequations}
\begin{align}
\la X(t_1) X(t_2) \ra_\text{ann} & \sim \left[t_1^{2\gamma} + t_2^{2\gamma} - |t_1-t_2|^{2\gamma}\right]\\
\la X(t_1) X(t_2) \ra_\text{det} & \sim  \left[(t_1+t_2)^{2\gamma} - |t_1-t_2|^{2\gamma}\right]
\end{align}
\end{subequations}
The two-time correlation function in the annealed case is exactly the same as for fractional Brownian motion with exponent $\gamma$.

The rest of the paper is organized as follows. In Sec.~\ref{sec:Riesz}, we define the Riesz gas. In particular, we use dimensional analysis to show that the behavior depends on two dimensionless numbers, the Riesz exponent $s$ and a  P\'{e}clet number which is essentially the ratio of typical interaction to noise. In Secs.~\ref{sec:Path-integral}--\ref{sec:2-time}, we focus on the  genuinely long-ranged Riesz gas with the exponent in the range $0<s<1$. In Sec.~\ref{sec:Path-integral}, we use a  path-integral formulation of the process and minimize an effective action to derive deterministic governing equations and boundary conditions, that play the role of the  MFT equations for our problem.  In Sec.~\ref{sec:PT}, we employ a perturbation approach and solve the governing equations in leading order. This allows us to determine the variances of the integrated current and the position of the tracer. Section~\ref{sec:2-time} is devoted to  two-time correlations. In Sec.~\ref{sec:concl}, we discuss our findings and outline possible future developments. In Appendix~\ref{app:SF}, we briefly consider effectively short-ranged Riesz gas ($s>1$). Details of derivations of the results of Sec.~\ref{sec:2-time} are relegated to Appendix~\ref{app:details}. 

\section{The Riesz gas}
\label{sec:Riesz}

We consider a gas of particles on the line interacting through the Riesz potential \eqref{Riesz-V} and undergoing independent Brownian motions with diffusion coefficient $D$. In the over-damped limit, the particle positions $x_i$ evolve according to coupled stochastic differential equations 
\begin{equation}
\label{eq:partseq}
\dot{x}_i =  g \sum_{j \neq i} \frac{x_i-x_j}{|x_i-x_j|^{2+s}} + \eta_i  
\end{equation}
The noise contributions $\eta_i$ are Gaussian with zero-mean and correlations
\begin{equation}
  \la \eta_i(t) \eta_j(t') \ra = 2 D \delta_{ij} \delta(t-t')
\end{equation}

To ensure that the gas with repulsive interactions does not freely expand, we consider a finite system with a large number $N$ of particles in a very shallow confining external potential. The density of particles in a large region around the origin is essentially uniform; we denote it by $\rho$. With such assumptions, our results for the tracer provide an intermediate asymptotic valid up to a crossover time $T_*(N)$ when the tracer eventually ``realizes" that the system is finite. The crossover time diverges as $N\to\infty$. In this limit, we can forget about the shallow confining potential and the finiteness of the system.

The uniform Riesz gas with $s>0$ is characterized by a single dimensionless parameter 
\begin{equation}
\label{def:G}
  G = \frac{g \rho^s}{D}
\end{equation}
This parameter $G$ measures the relative strength of interactions versus noise. Since $g \rho^{s+1}$ is a typical velocity of a particle caused by an adjacent particle and $\rho^{-1}$ is a typical distance between adjacent particles, $G$ plays a role of a P\'{e}clet number for the Riesz gas. The coupling constant $g$ and the diffusion coefficient $D$ have independent dimensions for $s>0$. One can use them to construct the units of length and time: $\left(\frac{g}{D}\right)^\frac{1}{s}$ and $\frac{1}{D}\left(\frac{g}{D}\right)^\frac{2}{s}$. Measuring length and time in terms of these units we can effectively set the coupling constant and diffusion coefficient to unity and take  $g=1=D$ in  the following. Then,  the problem only  depends  on the dimensionless density $\rho$, i.e., on $G^{1/s}$ in terms of the original variables. Note that  when $s=0$, i.e., for the Dyson gas, the coupling constant and diffusion coefficient have the same dimensions and can not be set to 1 independently.

The  coarse-grained density field  of the particles satisfies the continuity equation
\begin{equation}
\label{qJ:cont}
\p_t q  + \p_x J = 0
\end{equation}
where  $q = q(x,t)$ is the density and $J= J(x,t)$ is the local current. The current  $J$ contains the standard diffusion term, $-D\p_x  q= -\p_x  q$, plus another deterministic contribution $J_{{\rm Riesz}}$ arising from the Riesz potential  and a stochastic component due to the noise. Thus,  we  write
\begin{equation}
 \label{eq:defJRiesz}
 J = J_{{\rm Riesz}}  -  \p_x  q +  \sqrt{2 q}\, \eta
\end{equation}
The noise  $\eta=\eta(x,t)$ satisfies
\begin{equation}
  \la \eta(x,t)\ra =0, \quad   \la \eta(x,t) \eta(x',t') \ra = \delta(x-x') \delta(t-t')
\end{equation}
The amplitude of the noise, $\sqrt{2 q}$, reflects the Brownian nature of the point particles \cite{Spohn91,Kipnis99,Derrida07,MFT15}. 
The Riesz contribution  $J_{{\rm Riesz}}$  reads 
\begin{equation}
\label{eq:Jriesz}
J_{{\rm Riesz}} =
\begin{cases} 
   q  \Hil_{s}[q]                                &0<s<1 \\
  - (1+s) \zeta(s) q^{s} \p_x q      & s>1
\end{cases} 
\end{equation}
as we show below. In the $s>1$ range, this  Riesz current \eqref{eq:Jriesz} contains the zeta function $\zeta(s)$ (see Appendix~\ref{app:SF} for a derivation).
In the $0<s<1$ range, the Riesz current is expressed using  a modified Hilbert transform
\begin{equation}
\label{H:def}
\Hil_{s}[q] = \int d y\,  \frac{x-y}{|x-y|^{2+s}} \, q(y) 
\end{equation}
which reduces to the Hilbert transform in the $s\to 0$ limit. Hereinafter,  spatial integrals  over the entire line
will be denoted  $\int$, e.g., $\int dy\equiv  \int_{-\infty}^{\infty} dy$ in \eqref{H:def}. We also define  the potential $\Vil_{s}[q]$  at  $x$ due to the density profile $q$: 
\begin{equation}
\label{eq:Vildef}
\Vil_{s}[q] = \frac{1}{s} \int dy\, \frac{q(y)}  {|x-y|^{s}} 
\end{equation}
This potential satisfies $ \Hil_{s}[q] = -\p_x \Vil_{s}[q]$. The Riesz current \eqref{eq:Jriesz} can be derived from the general formula (expressing that the deterministic motion of particles is over-damped)
\begin{equation}
\label{Riesz-J}
J_{{\rm Riesz}}= - q\,\frac{\p}{\p x}\left(\frac{\delta}{\delta q} \mathcal{E}[q]\right) 
\end{equation}
where $\mathcal{E}$ is the interaction energy:
\begin{subequations}
\begin{equation}
\label{Feq:<}
\mathcal{E}[q] = \frac{1}{2 s} \int\int dx\,dy\,\,\frac{q(x) q(y)}{|x-y|^s} 
\end{equation}
when $0<s<1$, and
\begin{equation}
\label{Feq:>}
\mathcal{E}[q] = \frac{\zeta(s)}{s}\int dx\,  q^{1+s} 
\end{equation}
\end{subequations}
when $s>1$. The total deterministic current can be derived from 
\begin{equation}
\label{Riesz-D-J}
J_{{\rm Riesz}}  -  \p_x  q = - q\,\frac{\p}{\p x}\left(\frac{\delta}{\delta q} \mathcal{F}[q]\right) 
\end{equation}
with free energy
\begin{equation}
\label{Riesz-EF}
 \mathcal{F}[q] = \mathcal{E}[q] + \int dx\, q \ln q 
\end{equation}
that in addition to the interaction energy contains an entropic contribution. 

In the present work, we are primarily interested in Riesz gases with a long-ranged potential ($0<s<1$). For $s > 1$, the potential is effectively short-ranged and a well-understood single-file behavior emerges (as discussed in Sec.~\ref{sec:concl}). Our derivations rely on the fluctuating hydrodynamics together with its path-integral re-formulation, and the Martin-Siggia-Rose method \cite{MSR}. These tools are popular in macroscopic fluctuation theory, see \cite{Kurchan07,DG09,Kirone15a}. The Kawasaki-Dean method \cite{kawasaki94,dean96} can be alternatively used to derive the same expression \eqref{eq:Jriesz} for the Riesz contribution to the current.

\section{Hydrodynamics of the Riesz gas}
 \label{sec:Path-integral}

When $0<s<1$, the fluctuating hydrodynamics of one-dimensional stochastic Riesz gases is governed by the stochastic partial differential equation
\begin{equation}
\label{cont}
\p_t q = - \p_x\Big( q \Hil_{s}[q] - \p_x q + \sqrt{2 q}\,\eta \Big) 
\end{equation}
with $\Hil_{s}[q]$ given by \eqref{H:def}. Equation \eqref{cont} resembles the governing equation of fluctuating hydrodynamics of diffusive lattice gases \cite{Spohn91}. Analytical tools available  to investigate the statistical properties of lattice gases can be adapted to the present case to probe dynamical fluctuations in the Riesz gas. Namely, we shall develop a deterministic reformulation of fluctuating hydrodynamics analogous to the MFT of diffusive lattice gases \cite{Bertini01,Bertini02,Jordan03}. 

\subsection{Path-integral formalism}

The solution of the stochastic equation~\eqref{cont} can be expressed via a path integral. One writes the Gaussian measure for the white noise and integrates over it.  This procedure known as the Martin-Siggia-Rose method \cite{MSR} is standard; details can be found, for example, in the closely related context of the macroscopic fluctuation theory \cite{Kurchan07,DG09,Kirone15a}. The probability of transition from an initial configuration at $t =0$ to a final configuration at $t = T$  can be written as a functional integral after integrating out the white noise  $\eta(x,t)$:
\begin{equation}
\label{eq:Pfi}
P(q(x,T)|q(x,0)) = \int \int \int {\mathcal D}J\, {\mathcal D}q\, {\mathcal D}p\, e^{-\mathcal{S}}
\end{equation}
where
\begin{equation*}
\mathcal{S}=\int_0^T dt\int dx\,\left[\frac{(J - q \Hil_{s}[q]+  \p_x  q  )^2}{4 q} +  p (\p_t q + \p_x J)\right]
\end{equation*}
The second term in the integrand ensures that $q$ and $J$ obey the continuity equation \eqref{qJ:cont}, with $p=p(x,t)$ playing the role of the Lagrange multiplier. Evaluating the quadratic integral over $J$ yields 
\begin{equation}
\label{eq:Pfi2}
P(q(x,T)|q(x,0)) = \int \int {\mathcal D}q\, {\mathcal D}p\,
\,e^{- \int_0^T \int dt\,  dx\,   S(q,p) } 
\end{equation}
with action
\begin{equation}
\label{def:Action}
S(q,p) = p \p_t q -  q (\p_x p)^2 - q (\p_x p) \Hil_{s}[q]\,\p_x p + \p_x p\, \p_x q
\end{equation}
The form of \eqref{def:Action} is remarkably similar to the MFT action \cite{MFT15}. The new feature is the presence of $\Hil_{s}$ accounting for the long-ranged interactions.

\subsection{The cumulant generating function of an observable}

Take an arbitrary observable ${\mathcal O}(\{q(x,t),p(x,t)\})$. Its characteristic function can be written as 
\beq
\label{eq:pathcgfY}
\la e^{\lambda {\mathcal O}} \ra = \int \int  {\mathcal D}q\, {\mathcal D}p\,\, e^{\lambda {\mathcal O}- \int_0^T \int dt\,  dx\,   S(q,p) }\,P[q(x,0)]
\eeq
where $P[q(x,0)]$, the probability of the initial profile $q(x,0)$, represents how the  system is prepared at $t=0$.  For the deterministic initial condition with uniform  profile
$\rho$, we merely take
\begin{equation}
  P[q(x,0)] = \delta\left( q(x,0) - \rho \right)
\end{equation}
If the system is prepared with the equilibrium distribution of density profiles (annealed case), we take
\beq
P[q(x,0)] = \exp(-\mathcal{F}[q(x,0)])
\eeq
with free energy defined by \eqref{Feq:<} and \eqref{Riesz-EF}. The cumulant generating function (CGF)  is the logarithm of the characteristic function. The CGF encodes all cumulants $\la {\mathcal O}^n \ra_c$ of the observable  ${\mathcal O}$:
\beq
\label{def:CGF}
\mu(\lambda) = \ln{\la {\rm e}^{\lambda {\mathcal O}} \ra} = \sum_n \frac{1}{n!}
\la {\mathcal O}^n \ra_c
\eeq

In this work, we analyze two observables. The first is the integrated current $Q_T$ that has flown through the origin during the time interval $(0,T)$. It is  given by
\beq
Q(T)  = \int_{0}^{\infty} dx\,[q(x,T) - q(x,0)]
\label{def:Q_T}
\eeq
We note that $Q(T)$ depends only  on the final and the initial density profiles. 
Another observable is the position $X(T)$ of the tracer at time $T$; without loss of generality we set $X(0)=0$. The dynamics of the tracer is identical to the other particles, its tag allows us to focus on the same particle and thus study a self-diffusion phenomenon. In a single-file motion \cite{Richards,Pincus,Burlatsky,Diamant,BarkaiSilbey,Benichou13} particles cannot overtake each other, so the number of particles to the right of the tracer remains constant. We schematically write 
\begin{equation*}
\int_{X(T)}^\infty dx\, q(x,T) = \int_0^\infty dx\,q(x,0)
\end{equation*}
which in conjunction with \eqref{def:Q_T} gives \cite{Kirone14,SethuramanVaradhan2013,Imamura}
\begin{equation}
\label{eq:implicittr}
\int_0^{X(T)} dx   \, q(x,T) = Q(T)
\end{equation}
This useful relation implies that the statistics of $X(T)$ and $Q(T)$ are closely related. Below we first derive analytical results for the statistics of the current  and then translate them to the statistics of the position of the tracer.

\subsection{Governing equations and boundary conditions}

The action $S$ and the integrated current $Q(T)$  grow with time and in  the large-time limit, the path integral \eqref{eq:pathcgfY} will be  dominated by its saddle point \cite{Derrida07,MFT15}. The corresponding  optimal `trajectory'  $\{q(x,t),~p(x,t)\}$ is found by varying the action with respect to $q$ and $p$. For $0< t < T$, the Euler-Lagrange equations read
\begin{subequations}
\label{EL}
\beqa
(\p_t - \p_x^2) q &=& -\p_x \left( {2} q \p_x p + q \Hil_{s}[q] \right)  \label{eq:EL1} \\
(\p_t + \p_x^2) p &=& - (\p_x p)^2 -\Hil_{s}[q] \p_x p  + \Hil_{s}[q \p_x p]  \label{eq:EL2}
\eeqa
\end{subequations}
These equations differ from the equations of the macroscopic fluctuation theory \cite{Derrida07,MFT15} only by terms with $\Hil_{s}$.
Similar equations have also appeared in the study of the large $N$ limit of Harish-Chandra-Itzykson-Zuber integrals \cite{Matytsin,Instanton14}. 
The governing equations \eqref{EL} are usually universal, i.e., independent of the observable [see, however, Eq.~\eqref{eq:ELS2}], while the boundary conditions do  depend on the observable. In the  one-dimensional case when the observable is the integrated current, the saddle-point relations  at initial and final times  involve a contribution from $Q(T)$ \cite{DG09,Meerson12a,Kirone14}.
Hence, the boundary condition at $t =T$  reads
\beq
p(x,T) = \lambda \frac{\delta Q(T)}{\delta q(x,T)}  = \lambda \theta(x)
\label{Bcond:T}
\eeq
The initial condition at  $t=0$  depends on whether the initial preparation of the system is 
annealed or deterministic. We consider a system starting from a uniform density $\rho$, and hence, for the deterministic initial condition, we have
\begin{subequations}
\beq
q(x,0)|_{\text{det}} = \rho
\label{Cond:quench}
\eeq
In the annealed case, the system starts at equilibrium and density fluctuations are allowed. The free energy $\mathcal{F}$ is defined by 
\eqref{Feq:<} and \eqref{Riesz-EF}, and the initial condition for $p(x,0)$ reads
\beq
p(x,0)|_{\text{ann}} = -\lambda \frac{\delta Q(T)}{\delta q(x,0)} + \frac{\delta \mathcal{F}}{\delta q}\bigg\vert_{q(x,0)} 
\label{Cond:anneal}
\eeq
\end{subequations}
Thus we must solve Eqs.~\eqref{EL} subject to \eqref{Bcond:T} at the final time $T$ and the initial condition (\ref{Cond:quench}) in the deterministic case, or (\ref{Cond:anneal}) in the annealed case. The function $\mu(\lambda)$ is determined by substituting the solution  $\{q(x,t),~p(x,t)\}$ in the path integral \eqref{eq:pathcgfY}. This latter calculation can be significantly simplified by noting that
\begin{equation}
\label{formuleRahul}
\frac{d \mu}{d \lambda}  = \frac{\la {\mathcal O} {\rm e}^{\lambda {\mathcal O}}\ra}{ \la {\rm e}^{\lambda {\mathcal O} }  \ra }  
\end{equation}
Therefore the CGF is obtained by evaluating the value of the observable ${\mathcal O}$ for the optimal solution  $q(x,t),~p(x,t)$. This shortcut was noticed in Refs.~\cite{Stepanov01,Vivo16}, and efficiently used in a number of recent studies \cite{Meerson19,Meerson22a,Kirone22,Dandekar22,GrabschTracer}.

In the present case, we have 
\beq
\mu'(\lambda) = Q(T)
\label{eq:mup}
\eeq
with $Q(T)$ evaluated on the solution $q(x,t)$ of the governing equations \eqref{EL} with appropriate boundary conditions.

Handling a pair of non-linear, non-local coupled partial differential equations \eqref{EL} is mathematically daunting. These equations do not admit an analytical solution. Fortunately, a perturbative calculation based on expansion in $\lambda$ leads to {\em exact} results for the variance $\left\langle Q^2 \right\rangle$ as we show in the next section. To compute higher cumulants one needs higher orders in a perturbative expansion: The $(n+1)^\text{th}$ cumulant is determined by the solution of the MFT equations up to $n^\text{th}$ order in $\lambda$. At present, computing $\left\langle Q^4 \right\rangle_c$ seems analytically intractable.

\section{Perturbative solution}
\label{sec:PT}

We follow the strategy developed in \cite{Meerson12a} relying on the obvious fact that for $\lambda=0$, the solution follows the noiseless evolution, which in the present case is very simple:  $q(x,t)=\rho$ and $p(x,t)=0$ at all times. The expansion of $(q,p)$ in $\lambda$ generates the cumulants of the current. A  calculation at the lowest order enables one to determine the variance of $Q(T)$ both in the deterministic and the annealed ensembles.  
Up to the first order in $\lambda$ we have
\beqa
q = \rho + \lambda q_1 + O(\lambda^2), ~~~~ p = \lambda p_1 + O(\lambda^2) 
\label{eq:perts}
\eeqa
Plugging these expansions into Eqs.~\eqref{EL} we obtain 
\begin{subequations}
\label{EL-1}
\beqa
 \label{eq:EL11}
(\p_t - \p_x^2) q_1 &=&  - \rho \p_x \left(  \Hil_{s}[q_1] +   {2}\p_x p_1 \right)\\
(\p_t + \p_x^2) p_1 &=& \rho \Hil_{s}[\p_x p_1] 
\label{eq:EL21}
\eeqa
\end{subequations}
at  first order. (We have taken into account an obvious relation $\Hil_{s}[\rho] = 0$.) The boundary condition \eqref{Bcond:T} reads
\beq
p_1(x,T) = \theta(x) \label{eq:pfQ}
\eeq
The initial conditions \eqref{Cond:quench}--\eqref{Cond:anneal} become, at first order, 
\begin{subequations}
\begin{align}
 \label{eq:piQqu}
&q_1(x,0)|_{\text{det}} = 0    \\
\label{eq:piQan}
&p_1(x,0)|_{\text{ann}}   = \theta(x) +
\frac{\delta \mathcal{F}}{\delta q}\bigg\vert_{q_1(x,0)}
\end{align}
\end{subequations}

\subsection{Variance of the integrated current: Deterministic case}
\label{subsec:Var-Q-det}

The first-order equations \eqref{EL-1} can be solved via Fourier transform
\beq
\widehat{f}(k) = \int dx\, {\rm e}^{-\ii\,k x}  f(x)  dx
\eeq
(Recall the short notation $\int dx\equiv \int_{-\infty}^\infty dx$ for the spatial integrals over entire line.) A very useful identity 
\beq
\widehat{\Hil_{s}[f]}(k) = - \frac{\ii\, \sqrt{\pi}}{2^s}\,
 \frac{\Gamma\left(\frac{1-s}{2}\right)}{\Gamma\left(1 +\frac{s}{2}\right)}\,\,
 k\, |k|^{s-1} \widehat{f}(k) 
\eeq
follows from the general formula \cite{Gelfand,RieszGF}
\beqa
\int dx\, {\rm e}^{-\ii\, k x} |x|^{b} 
= \frac{2^{b +1}\sqrt{\pi}}{|k|^{b +1}}\, \frac{\Gamma \left( \frac{1+b}{2}\right) }
      {\Gamma\left(-\frac{b}{2}\right)}
 \label{Gelf}
 \eeqa
Equation~(\ref{eq:EL21}) becomes
\begin{equation}
\label{p1:eq}
  \p_t \widehat{p}_1 =  \omega(k) \widehat{p}_1
\end{equation}
with dispersion relation 
\begin{subequations}
\begin{align}
\label{def:omegak}
 & \omega(k) =   k^2 +  A_s |k|^{s+1} \\
 \label{def:As}
 & A_s = \frac{ \rho \sqrt{\pi}}{2^s}\frac{\Gamma\left(\frac{1-s}{2}\right)}{\Gamma\left(1 +\frac{s}{2}\right)}
\end{align}
\end{subequations}
The boundary condition \eqref{eq:pfQ} gives $\widehat{p_1}(k,T) = \frac{1}{ \ii\, k}$,  and hence \eqref{p1:eq} leads to 
\beq
\widehat{p}_1(k,t) = \frac{1}{\ii\,k}\, e^{-  \omega(k) (T-t)}
\label{eq:Ca}
\eeq

The equation for $q_1$ is solved along similar lines. The Fourier transform  of \eqref{eq:EL11} is 
\beq
[\p_t + \omega(k)]\widehat{q}_1 =   2 \rho k^2 \widehat{p}_1
\label{eq:FTq1}
\eeq
The initial condition~(\ref{eq:piQqu}) gives  $\widehat{q}_1(k,0) =0$. Solving \eqref{eq:FTq1} with $ \widehat{p}_1$ given by  \eqref{eq:Ca} we obtain 
\beq
\widehat{q}_1(k,t) = \rho k\, 
\frac{e^{-\omega(k)(T-t)}  -  e^{-\omega(k)(T+t)}}{\ii\,\omega(k)}
\label{solq1Quench}
\eeq

Using \eqref{formuleRahul} we find the cumulant generating function at lowest  order in $\lambda$:
\begin{eqnarray}
   \mu'(\lambda) = \lambda \int_0^\infty dx\,\left[q_1(x,T) - q_1(x,0)\right] 
 \end{eqnarray}
In Fourier  space, this gives
\begin{eqnarray}
 \mu'(\lambda) & = & \ii\, \lambda \int_{-\infty}^\infty \frac{\widehat{q}_1(k,T)-\widehat{q}_1(k,0)}{k} \frac{dk}{2 \pi} \nonumber \\
                        & =  &\lambda \rho  \int_{-\infty}^\infty  \frac{1 - {\rm e}^{-2 \omega(k)T}}{\omega(k)} \frac{dk}{2 \pi}
  \label{muprimeqFT}
\end{eqnarray}
The asymptotic $T \to \infty$ behavior of the above integral is dominated by the $|k|^{1+s}$ term as  the diffusive part $k^2$ in $\omega(k)$ becomes  irrelevant (as readily  seen by redefining  $\kappa :=  k T^{\frac{1}{s+1}}$). Thus, in the large time limit, we find
\begin{equation*}
 \int_{-\infty}^\infty
    \frac{1 - {\rm e}^{-2 \omega(k)T}}{\omega(k)} \frac{dk}{2 \pi}
    \to  \frac{T^{\frac{s}{s+1}}}{\pi}  \int_{0}^\infty d\kappa\,
    \frac{1 - {\rm e}^{-2 A_s \kappa^{s+1}}}{ A_s \kappa^{s+1}}
\end{equation*}
The second integral can be computed \cite{Abram} leading to
\begin{equation}
   \left\langle Q^2 \right\rangle_{{\rm det}} = W_s (2  T)^{\frac{s}{s+1}}, \qquad 
   W_s = \frac{\rho\, \Gamma\big(\frac{1}{s+1}\big)}{\pi s A_s^{\frac{1}{s+1}}}
\label{VarQquench}
\end{equation}
where we have taken into account that, by definition of the CGF in \eqref{def:CGF}, the first order term in  $\mu'(\lambda)$ represents the variance of the current. Recalling an explicit formula \eqref{def:As} for $A_s$, we
write \eqref{VarQquench} as  
\begin{equation}
\label{Qs-det}
 \langle Q^2 \rangle_{{\rm det}} =  (\rho T)^\frac{s}{s+1} U_s
\end{equation}
with amplitude
\begin{equation}
\label{Us-det}
U_s = \frac{\Gamma\big(\frac{1}{s+1}\big)}{s }
 \left[\frac{4^s\,\Gamma\left(1 +\frac{s}{2}\right)}{
 \pi^{s+3/2}\,\Gamma\left(\frac{1-s}{2}\right)}\right]^\frac{1}{s+1}
\end{equation}
depending only on the Riesz exponent $s$. 

The variance of the current across the origin increases as $T^{s/(s+1)}$, i.e., slower than for single-file diffusive systems \cite{Arratia83,KRB,Kirone15a} where the exponent is $1/2$. The exponent $\frac{s}{s+1}$ approaches $\frac{1}{2}$  when  $s \uparrow 1$, albeit the amplitude $U_s$ vanishes in this limit, $U_s\to \sqrt{(1-s)/\pi}$. This indicates that precisely at $s=1$, the growth of the variance might be slower than $\sqrt{T}$, possibly  with a logarithmic correction (see Sec.~\ref{sec:concl}). 

Formulae \eqref{Qs-det}--\eqref{Us-det} are also singular when $s\downarrow 0$, although the hydrodynamic equations remain well-defined. This may be an indication that the function $\mu(\lambda)$ is itself singular for $s \rightarrow 0$ and that the perturbative scheme breaks down in this limit. 

When $s<1$, the diffusive contribution is subdominant in the long time limit. This could have been anticipated by observing that the second order derivatives in Eqs.~\eqref{EL} are negligible in the scaling limit compared to the Hilbert operator $\Hil_{s}$. Physically, this means that, in the limit we consider, the Riesz current is dominated by the advection term coming from the  interactions, rather than by the diffusive flux of entropic origin.

\subsection{Variance of the integrated current: Annealed case}

In the annealed case, the boundary condition at the final time $T$ is the same as in the deterministic case, so Eq.~\eqref{eq:Ca} still holds. To implement the initial condition \eqref{eq:piQan}, we need an appropriate expression for the free energy. For $0<s<1$, the expression \eqref{Feq:<} is schematic, e.g., it diverges. To avoid the divergence, we subtract $\rho$ from $q(x)$ and $q(y)$ in \eqref{Feq:<}, and also subtract a constant from the entropic contribution so that it vanishes at infinity. This gives
\begin{eqnarray}
\label{Free-E}
\mathcal{F}[q] &=& \int dx\, q(x)\, \ln \frac{q(x)}{\rho} \nonumber \\
&+&  \int \int dx dy\,\,\frac{(q(x)-\rho) (q(y)-\rho)}{2s\, |x-y|^s}
\end{eqnarray}
Plugging \eqref{Free-E} into \eqref{eq:piQan} together with expansion \eqref{eq:perts} we obtain 
\beq
p_1(x,0) =  \theta(x)  + \frac{1}{s} \int dy\,
\frac{ q_1(y,0)}{|x-y|^{s}} +  \frac{q_1(x,0)}{\rho}
\eeq
in the first order. Performing the Fourier transform of this relation and using Eqs.~\eqref{Gelf},  \eqref{def:omegak},  and \eqref{eq:Ca},
 we derive  the initial value $\widehat{q}_1(k,0)$
in the annealed case 
\beq
\label{q1-k:IC}
\widehat{q}_1(k,0) =   \frac{  \rho k}{\ii\, \omega(k) }
\left( {\rm e}^{-  \omega(k)T} -1 \right) 
\eeq
Equation \eqref{eq:FTq1} is still valid, but now we have to solve it subject to the 
 initial condition \eqref{q1-k:IC}. The solution reads 
\beq
\widehat{q}_1(k,t) =  \rho k\,
\frac{e^{-\omega(k)(T-t)}  -  e^{-\omega(k)t}}{\ii\, \omega(k)} \label{solq1Ann}
\eeq

To establish the cumulant generating function in the lowest order we proceed as before and find 
\begin{eqnarray}
\label{mu-annealed}
 \mu'(\lambda) & = & \ii\, \lambda \int_{-\infty}^\infty \frac{\widehat{q}_1(k,T)-\widehat{q}_1(k,0)}{k} \frac{dk}{2 \pi} \nonumber \\
                        & =  &\lambda \rho  \int_{-\infty}^\infty  \frac{1 - e^{-\omega(k)T}}{\omega(k)} \frac{dk}{ \pi}
\end{eqnarray}
The long time behavior is extracted similarly to the deterministic case (cf. equation \eqref{VarQquench}). We finally arrive at a simple relation
 between the variances  in the annealed and deterministic cases:
\beqa
\la Q^2 \ra_{\rm{ann}} &=& 2^\frac{1}{1+s}\, \la Q^2 \ra_{\rm{det}}
\eeqa
As expected, the variance in the annealed case is enhanced compared to the deterministic setting because of the non-zero initial fluctuations. This result is another example of the ever-lasting influence of initial conditions \cite{Barkai}. The ratio, $2^{\frac{1}{1+s}}$, approaches as $s \rightarrow 1$ the value $\sqrt{2}$ found in hard-core single-file systems such as the symmetric exclusion process \cite{Meerson12a}.

\subsection{Variance of the position of the tracer}

The position of the tracer can be determined by the fact that particles in the Riesz gas with $s\geq 0$ can not overtake one another. The motion of the tracer displaces the particles ahead of it and drives a current through the system. At first order in $\lambda$, we have from \eqref{eq:implicittr},
\begin{equation}
\label{X-Q}
X(T) = \frac{\lambda}{\rho} \int_0^{\infty} dx\, [q_1(x,T) - q_1(x,0)]
\end{equation}
which differs from the integrated current only by a $1/\rho$ factor. This simple relation is valid only at the  first order. At higher orders in $\lambda$, 
the number of particles in the vicinity of the tracer is random and its statistics must also be taken into account \cite{DG09,Kirone15a,Imamura,Benichou22,Benichou23}. 

Using \eqref{X-Q} we find
\beq
\la X^2 \ra = \frac{1}{\rho^2}\, \la Q^2 \ra
\eeq
in the leading order. This is valid both for the deterministic and annealed cases. Therefore 
\beqa
\la X^2 \ra_{\rm{ann}} = 
2^{\frac{1}{1+s}}\, \la X^2 \ra_{\rm{det}}
\eeqa
The variance of the position of the tracer scales as a fractional Brownian motion with Hurst exponent $\gamma = \frac{1}{2}\frac{s}{1+s}$ \cite{Mandelbrot,Meerson22}.  The vanishing of the exponent as $s \rightarrow 0$ is consistent with the logarithmic mean square displacement of a tracer in Dyson's model of interacting Brownian particles established by Spohn \cite{Spohn87, Spohn87b}. Note that for the symmetric exclusion process, a lattice gas with hard-core interacting particles (that heuristically corresponds to $s \to \infty$), the statistical identity between the tracer process and the fractional Brownian motion with exponent $\frac{1}{4}$ has been proved in \cite{Sundar}.

\section{Two-time correlations}
\label{sec:2-time}

In this section, we investigate the two-time correlations of the process. The results  provide further indication that a tracer in the Riesz gas behaves as a fractional Brownian motion. The saddle-point method used to determine quadratic fluctuations can be extended to unequal time correlations by introducing a source term in the optimal equations. This  will allow us to calculate two-time correlations of the integrated current and the tracer's position.

\subsection{Generating functional}

To derive current-current correlations at different times, we introduce the generating functional (see \cite{Derrida07,Kirone15b} for a detailed presentation of the formalism):
\begin{eqnarray}
\label{eq:genfunccorr1}
Z[\lambda(t)] &=&  \left\la {\rm exp}\left[\int_0^T dt\,  \lambda(t) Q(t)\right] \right\ra \nonumber\\
&=&  \left\la e^{\int_0^T dt\int  dx
 \lambda(t)    \theta(x)  (q(x,t)- q(x,0))} \right\ra
\end{eqnarray}
Two-time correlation function  of the current at times $t_1,t_2 < T$ 
can be found by taking functional derivatives of this  generating functional 
\beq
 \mu[\lambda(t)] =  \ln Z[\lambda(t)]
 \eeq
For example, recalling that $\langle Q(t) \rangle = 0$ for all $t$, we can expand the generating functional at lowest order with respect to the source-function $\lambda(t)$:
\beq
\mu[\lambda(t)] =  \int_0^T dt \int_0^T dt'   \lambda(t)  \lambda(t')
C(t,t') + \ldots
\label{FunctCGF}
 \eeq  
where $C(t,t') = \la Q(t) Q(t') \ra_c$ is the two-time correlation function. Higher order terms generate multiple-time correlation functions. Writing the average as a functional integral as in equation~\eqref{eq:Pfi2}, we observe that the bulk action $S(q,p)$  given in~\eqref{def:Action} is tilted  by the source term, $S(q,p) \rightarrow S_{\lambda(t)}(q,p)$, with 
\beq
S_{\lambda(t)}(q,p) = S(q,p) -  \lambda(t)    \theta(x)  (q(x,t)- q(x,0))
\label{def:Action2}
\eeq
Taking the functional derivative of $\mu[\lambda(t)]$  leads to
\beq
\frac{\delta \mu[\lambda(t)] }{\delta \lambda(t_1)} = \frac{\la Q(t_1) \exp(\int_0^T \lambda(t) Q(t)) \ra}{\la \exp(\int_0^T \lambda(t) Q(t)) \ra} = \la Q(t_1) \ra_{[\lambda]}
\eeq
where the final average is over the tilted action $S_{\lambda(t)}$. Comparing with \eqref{FunctCGF}, we deduce that the two-time correlations are given by 
\beq
C(t_1,t_2) = \frac{\delta \la Q(t_1) \ra_{[\lambda]} }{\delta \lambda(t_2)}
\Biggr|_{\lambda \equiv 0}
\label{CorrDQ}
\eeq
Thus, it suffices to determine the average with respect to the tilted action to access the correlations. As in the previous sections, we determine the generating functional by writing the Euler-Lagrange equations and solving them perturbatively at the lowest order.

\subsection{The Euler-Lagrange equations with a source}

In presence of the source term, the tilted action $S_{\lambda(t)}$ affects only the equation for $p$
\begin{eqnarray}
 \label{eq:ELS2}
(\p_t + \p_x^2) p &=& - \lambda(t) \theta(x)  - (\p_x p)^2 \nonumber\\
&-& \Hil_{s}[q] \p_x p  + \Hil_{s}[q \p_x p] 
\end{eqnarray}
while $q$ satisfies the same equation \eqref{eq:EL1} as before. 
At final time, we have
\beq
p(x,T) = 0
\label{BC:Tsource}
\eeq
The initial conditions depend on the setting:
\begin{subequations}
\begin{align}
\label{q-IC:det}
&q(x,0)|_{\text{det}} = \rho \\
\label{p-IC:anealed}
&p(x,0)|_{\text{ann}} =\theta(x)\int_0^T dt \lambda(t)  + \frac{\delta \mathcal{F}}{\delta q}\bigg\vert_{q(x,0)} 
\end{align}
\end{subequations}

We write $q = \rho + q_1$ and $p = p_1$, where $q_1$ and $p_1$ are linear functionals of  $\lambda(t)$. Performing a perturbative expansion we obtain
\begin{subequations}
\beqa
(\p_t - \p_x^2)q_1 &=& - \rho \p_x( \Hil_{s}[q_1] + {2} \p_x p_1) \label{eq:q1source} \\
(\p_t + \p_x^2)p_1 &=& \rho \Hil_{s}[\p_x p_1] -\lambda(t) \theta(x)  \label{eq:p1source} 
\eeqa
\end{subequations}
in the first order. The Fourier transform of $p_1$ reads 
\beqa
 \widehat{p}_1(k, t) 
 = \frac{1}{\ii\,k}  \int_t^T d\tau \lambda(\tau) {\rm e}^{\omega(k) (t-\tau)}
 \label{p1source}
\eeqa
with $\omega(k)$ defined in Eqs.~\eqref{def:omegak}--\eqref{def:As}. 

We begin with the deterministic setting. In this case $q_1(x,0)=0$ and the solution of \eqref{eq:q1source} reads
\beq
\widehat{q}_1(k, t) =  - 2 \ii\,\rho k 
\int_0^t d\tau \int_\tau^T dt_2 \lambda(t_2){\rm e}^{\omega(k)(2 \tau - t_2 -t)}
\label{eq:q1corrQuen}
\eeq
To determine the two-time correlation function we write
\begin{eqnarray}
\langle Q(t_1) \rangle  &=&
\int_0^\infty dx \left( {q}_1(x,t_1)- {q}_1(x,0) \right) \nonumber \\
 &=& \ii  \int_{-\infty}^\infty  \frac{\widehat{q}_1(k,t_1) - \widehat{q}_1(k,0) }{k}
\frac{dk}{2 \pi} \nonumber \\
&=&  \int_0^T dt_2 \,   C(t_1,t_2)\lambda(t_2)
 \label{eq:Q1avQuen}
\end{eqnarray}
In the last step we have used \eqref{CorrDQ} at the first order. Substituting $\widehat{q}_1$ and exchanging the order of the integrals (see Appendix~\ref{app:details} for details), we determine the two-time correlation function. In the large time limit, this correlation function has a neat form
\begin{equation}
\label{CorrelQu}
    C_{{\rm det}}(t_1,t_2) = W_s \left\{(t_1 + t_2)^{\frac{s}{1+s}} - |t_1 - t_2|^{\frac{s}{1+s}} \right\}
\end{equation}
with $W_s$ defined in Eq.~\eqref{VarQquench}.

In the annealed  setting, the initial condition \eqref{p-IC:anealed} leads to 
\begin{equation}
\label{init:anneal}
 \widehat{q}_1(k,0) = \frac{  \rho k}{\ii\, \omega(k) }  \int_0^T d\tau \lambda(\tau) ({\rm e}^{-\omega(k)\tau} - 1)
\end{equation}
in the  first order. Using this we determine $\widehat{q}_1(k,t)$ and use Eq.~\eqref{eq:Q1avQuen} to calculate the generating functional at lowest order (see Appendix~\ref{app:details}). In the large time limit, the two-time correlation function has again a neat form
\begin{equation}
\label{CorrelAnn}
C_{{\rm ann}}(t_1,t_2)= W_s
\left(t_1^{\frac{s}{1+s}} + t_2^{\frac{s}{1+s}} - |t_1-t_2|^\frac{s}{1+s} \right)
\end{equation}
As was explained in \cite{Kirone15b,Krug}, the results in the annealed case can be deduced from the deterministic case by letting the system
evolve up to a time $t_0$ and then measuring the  current $Q_{+}(t) = Q(t_0 +t) - Q(t_0)$ that has flown from that time on. Then, if we  calculate, with the help of \eqref{CorrelQu}, the deterministic correlations for $Q_{+}$ at times $t_1,t_2$,  and assume that $ t_0 \gg t_1,t_2$, we obtain \eqref{CorrelAnn}. Indeed, by shifting the time by the large duration $t_0$, the system is effectively put into equilibrium. 

For the tracer position, the two-time correlations are obtained by multiplying the two-time current-current  correlation function by the factor $\rho^{-2}$. We observe that these expressions are the same as those for a fractional Brownian motion with Hurst exponent $H=\frac{s}{1+s}$ \cite{Mandelbrot}. This further suggests that the tracer behaves as a fractional Brownian process.

\section{Discussion}
\label{sec:concl}

We studied current and tracer fluctuations in the one-dimensional Riesz gas. We focused has been on the genuinely long-range interaction regime, $0<s<1$, for which the collective dynamics  differs significantly from usual single-file systems. We derived integro-differential MFT-type equations, with non-local terms, that in principle allow one to probe large deviations in one-dimensional stochastic Riesz gases. Using these equations we investigated fluctuations of the integrated current $Q$ and the position $X$ of the tracer in the one-dimensional stochastic Riesz gas. By applying a perturbation approach to governing MFT-type equations we established the variances $\langle Q^2 \rangle$ and  $\langle X^2 \rangle$. The calculation of higher cumulants, e.g., $\langle Q^4 \rangle$ and $\langle X^4 \rangle$, remains an analytical challenge. 

We now restate some of our main results and outline potential extensions in terms of the original variables so that the dependence on the coupling constant $g$ and diffusion constant $D$  will be visible. For instance, in the genuinely long-range regime, $0<s<1$, the variance of the position of the tracer grows as
\begin{equation}
\label{Xs-det}
 \langle X^2 \rangle_{{\rm det}} = U_s G^{-\frac{1}{s+1}} \rho^{-2}(\rho^2 D T)^\frac{s}{s+1}
\end{equation}
in the deterministic case; in the annealed case, the variance is larger by a factor $2^\frac{1}{s+1}$. The growth law \eqref{Xs-det} depends on the dimensionless parameter $G=g\rho^s/D$ measuring the relative strength of interactions versus noise and  the amplitude $U_s$ is given by \eqref{Us-det}.

We begin with the short-range regime, $s>1$, when the extension of known results is rather straightforward once the transport coefficients are established.

\subsection{The short-range regime $s>1$}

The short-range regime $s>1$ can be dealt with by borrowing known results for the single file-diffusion derived from standard MFT  formalism \cite{Kirone14,Kirone15a}. When $s >1$, we observe from \eqref{Feq:>}, \eqref{Riesz-D-J}--\eqref{Riesz-EF} that  the deterministic current has the form  
\begin{equation}
\label{JD:+1}
J=-D(\rho)\p_x \rho, \quad  D(\rho) = D+(1+s) \zeta(s) g \rho^s
\end{equation}
The mobility of particles undergoing Brownian motions is $\sigma(\rho)=2D\rho.$  Thus for $s>1$, the governing equations are bona fide MFT equations and  we can use the general formula \cite{Kirone15a} for the self-diffusion of a tracer 
\begin{equation}
\la X^2 \ra|_\text{ann} = \sqrt{2}\,\la X^2 \ra|_\text{det} = \frac{\sigma(\rho)}{\rho^2}\sqrt{\frac{T}{\pi D(\rho)}}
\end{equation}
in single-file hydrodynamics characterized by transport coefficients $D(\rho))$ and $\sigma(\rho)$. By specializing to the short-range Riesz gas, we deduce 
\begin{equation}
\label{Xs-det-plus}
\la X^2 \ra|_\text{det} = \frac{1}{\sqrt{1+(1+s)\zeta(s)G}}\,\sqrt{\frac{2 D T}{\pi\,\rho^2}}
\end{equation}
In the annealed case, the variance is $\sqrt{2}$ times larger. When the relative strength of interaction vanishes, i.e.,  $G\to 0$, we recover the well-known behavior for Brownian particles undergoing single-file diffusion. The $T^{1/2}$ temporal growth remains the same independently of $G$, while the amplitude decays as $G$ increases. 

Correlation profiles in the frame of the tracer have been recently established \cite{Benichou21,Benichou22,Benichou23} for several single-file systems. We mention one neat formula applicable to interacting point particles undergoing independent identical Brownian motions and satisfying the single-file constraint. In these gases, the density $n(x+X,t)$ on distance $x$ from the tracer is correlated with the position $X$ of the tracer according to \cite{Benichou21}
\begin{equation}
\label{xXt}
\langle n(x+X, t)\, X\rangle|_\text{ann}  = \frac{\text{sign}(x)}{2D(\rho)/D}\,\text{Erfc}\left(\frac{|x|}{\sqrt{4D(\rho)t}}\right)
\end{equation}
in the annealed case. Specializing this formula to $D(\rho)$ given by \eqref{JD:+1} we obtain the correlation function \eqref{xXt} for the stochastic Riesz gas in the short-range regime.

\subsection{The case $s=1$}

The marginal $s=1$ case separates long-range and short-range regimes. The Riesz gas with $s=1$ corresponds to a  physically relevant system of particles confined to a one-dimensional line and interacting through the three-dimensional Coulomb potential. Therefore the Riesz gas with $s=1$ deserves a separate careful investigation. Here we use heuristic arguments to guess a plausible asymptotic behavior for the tracer's fluctuations. When $ s \to 1$, the  $s-$dependent term in \eqref{Xs-det-plus} diverges as 
\begin{equation}
\label{est}
1+(1+s)\zeta(s)\,\frac{g\rho^s}{D} \to 2\,\frac{g\rho}{D}\,\frac{1}{s-1}
\end{equation}
where we have used the asymptotic $\zeta(s)\simeq (s-1)^{-1}$ of the zeta function $\zeta(s)$ near $s=1$. The characteristic dimensionless diffusive length scale is $\ell \sim \sqrt{\rho^2 DT}$, and as long as $\ell^{-1}\sim \ell^{-s}$, there is no difference between the Riesz gas with exponent $s$ and the Coulomb gas with $s=1$. Therefore $(\rho^2 DT)^{s-1}\sim 1$, from which we deduce $(s-1)^{-1}\sim \ln(\rho^2 DT)$, and \eqref{est} becomes
\begin{equation}
\label{est-2}
1+(1+s)\zeta(s)\,\frac{g\rho^s}{D} \to \, \ln(\rho^2 DT), \qquad G=\frac{g\rho}{D}
\end{equation}
Plugging this into  \eqref{Xs-det-plus} yields
\begin{equation}
\label{Var-Coulomb}
\la X^2 \ra \sim \frac{1}{\sqrt{G}}\,\,\sqrt{\frac{D T}{\rho^2 \, \ln(\rho^2 DT)}}
\end{equation}
The dependence of the variance \eqref{Var-Coulomb} on the P\'{e}clet number $G$ is natural given the behavior in the $s<1$ and $s>1$ regimes, \eqref{Xs-det} and \eqref{Xs-det-plus}. Re-writing \eqref{Var-Coulomb} as
\begin{equation}
\label{Var-Coulomb-rho}
\la X^2 \ra \sim \rho^{-3/2}\,\,\sqrt{\frac{D^2 T}{g\,\ln(\rho^2 DT)}}
\end{equation}
emphasizes the $\rho^{-3/2}$ dependence on the density. 

\subsection{Riesz gases with $s\leq 0$}

The one-dimensional Riesz gas with $s=0$, more precisely a log-gas known in one dimension as the Dyson gas \cite{Dyson62}, satisfies the single-file constraint (i.e., particles cannot overcome each other). The exponent $\gamma=\frac{s}{2(s+1)}$ vanishes as $s\to +0$. Hence one anticipates a slower than algebraic growth, and indeed the variance increases logarithmically with time \cite{Spohn87}. It would be interesting to extend this prediction, e.g., to establish the dependence on the P\'{e}clet number, $G=g/D$ for the Dyson gas. 

When $s<0$, the overcome becomes feasible, so the tracer apparently exhibits a standard diffusive behavior:
\begin{equation}
\la X^2 \ra = 2\,F_1(G,s) DT, \qquad s < 0
\end{equation}
The case of $s=-1$ is particularly intriguing as it corresponds to the particles interacting through the one-dimensional Coulomb potential. 

Our computations of the position $X$ of the tracer in the one-dimensional Riesz gas with $s>0$ relied on the connection of $X$ with $Q$. This crucial feature is lost for Riesz gases with $s<0$. It is perhaps feasible to extend our techniques and compute the variance of the current, $\la Q^2 \ra$, for Riesz gases with $s<0$. New ideas and techniques are needed to probe $\la X^2 \ra$. 

\subsection{Higher dimensions}

The most interesting challenge is to  extend our analysis to stochastic Riesz gases in higher dimensions ($d \ge 2$). As the single-file phenomenon is absent for $d >1$, the problem seems simpler at first sight. Also, the MFT framework admits a straightforward extension (with non-local terms if $s <d$). However, our approach based on the relation between the position and the current is no longer applicable and new ideas are required.

We anticipate that the tracer behaves diffusively in the short-range $s>d$ regime:
\begin{equation}
\la {\bf R}^2 \ra = 2d\,F_d(G,s) DT, \qquad s>d
\end{equation}
Thus  $F_d(G, s) D$ is the self-diffusion coefficient. The challenge is to compute $F_d(G, s)$ as a function of the P\'{e}clet number $G=g\rho^{s/d}/D$ and the Riesz exponent $s$. 

The derivation of the exact formula for $F(G)$ looks like an unattainable goal. (Indeed, for the simple exclusion process, the self-diffusion coefficient is unknown already on the square lattice.) Perhaps, one can probe the asymptotic behavior of $F(G)$ in the large $G$ limit. In the case of  vanishingly small  $G$, we have non-interacting Brownian particles, so $F(0)=1$. In the long-range regime, $s<d$, a sub-diffusive behavior is expected,  $\la {\bf R}^2 \ra\sim T^{\beta(s,d)}$, with an unknown exponent  $\beta(s,d)<1$ when $s<d$. 

In two dimensions, the Riesz gases with $s=0, 1, 2$ are particularly interesting. We  expect sub-diffusive behaviors for the Ginibre gas ($s=0$)  and the Coulomb gas ($s=1$). The Calogero gas (usually studied \cite{Calogero69,Calogero71,Alexios06} in one dimension) is marginal in two dimensions, so similar to \eqref{Var-Coulomb} logarithmic corrections are plausible, such as $\la {\bf R}^2 \ra \sim DT/\ln(\rho DT)$. 

The self-diffusion phenomenon in the Riesz gases in dimensions $d\geq 2$ appears intractable with available tools. The MFT framework for high-dimensional Riesz gases could be applied, however, to more tractable problems such as void formation \cite{Meerson12b}.

\bigskip
\noindent{\bf Acknowledgments.}
We thank H. Spohn for an inspiring discussion and S. Mallick for a careful reading of the manuscript. PLK thanks IPhT Paris-Saclay for excellent working conditions. The work of KM has been supported by the project RETENU ANR-20-CE40-0005-01 of the French National Research Agency (ANR).

\appendix
\section{Interaction energy when  $s \geq 1$}
\label{app:SF}

Despite the simplicity and beauty of Eq.~\eqref{Feq:>}, its derivation is long and far from rigorous. We refer to \cite{Kundu19} for derivation of a similar result in the case of harmonically confined particles. Here we limit ourselves to a few no-rigorous arguments in favor of \eqref{Feq:>}. 

First, we notice that Eq.~\eqref{Feq:<} giving the interaction energy in the $s<1$ regime is a natural continuous version of the exact formula 
\begin{equation}
\label{Riesz-E}
\mathcal{E} = (2s)^{-1} \sum_{i\ne j} \frac{1}{|x_i-x_j|^s}
\end{equation}
(Every $i\ne j$ in \eqref{Riesz-E} appears twice, hence the factor $1/2$.)

A singularity at $x=y$ in the integral in Eq.~\eqref{Feq:<} is integrable if $s<1$, so we can use the integral representation \eqref{Feq:<} of the sum \eqref{Riesz-E}. The integral still diverges in an infinite system, but we put it under the carpet. 

When $s>1$, the singularity at $x=y$ in the integral in Eq.~\eqref{Feq:<} leads to divergence. We thus return to summation and compute the energy per particle by fixing $i$ and summing over all $j\ne i$, or equivalently over $n=j-i$
\begin{equation}
\label{E-particle}
\mathfrak{e} = (2s)^{-1}\sum_{n\ne 0}\frac{1}{|n/q(x)|^s}=s^{-1}\zeta(s) q^{s}
\end{equation}
We have relied on a crucial {\em assumption} that the spatial distribution of particles is locally equidistant; see \cite{Kundu19} for justification. This assumption has allowed us to write $x_j-x_i=n/q(x)$ in \eqref{E-particle}, where $x$ means $x_i$. The total energy is 
\begin{equation}
\label{E-zeta}
\mathcal{E}[q] = \int dx\,q\mathfrak{e}= s^{-1}\zeta(s) \int dx\,q^{s+1}
\end{equation}
with extra $q$ in the first integral since $\sum_i\to \int dx\,q(x)$. Equation \eqref{E-zeta} is the announced Eq.~\eqref{Feq:>}. 

When $s=1$, the sum in \eqref{E-particle} diverges. It seems reasonable to use the diffusive scale $\sqrt{T}$ as an upper cutoff in the sum. Thus \eqref{E-particle} gives 
\begin{equation}
\label{E-particle-1}
\mathfrak{e} = \sum_{n=1}^{\sqrt{T}}\frac{q(x)}{n}=q(x)\, \ln \sqrt{T} = \tfrac{1}{2}q \ln T
\end{equation}
Therefore 
\begin{equation}
\label{E-1}
E[q] = \int dx\,q\mathfrak{e} = \tfrac{1}{2}\ln T \int dx\,q^2
\end{equation}
and in the long-time limit $D(q) = q\ln T$. These heuristic arguments give
\begin{equation}
\label{Var-1}
\la X^2 \ra|_\text{ann} = \sqrt{2}\,\la X^2 \ra|_\text{det} = \rho^{-3/2}\,\sqrt{\frac{4T}{\pi\ln T}}
\end{equation}
In Sec.~\ref{sec:concl}, we presented heuristic arguments leading to Eq.~\eqref{Var-Coulomb-rho} for the variance. This result is consistent with Eq.~\eqref{Var-1}. The only difference is that Eq.~\eqref{Var-Coulomb-rho} is agnostic to the numerical pre-factor. (If in the estimate \eqref{E-particle-1} we take the scale $T^{1/4}$ of a typical displacement of the tracer as an upper cutoff, this would enhance \eqref{Var-1} by a factor  $\sqrt{2}$.)  A more careful treatment of the $s = 1$ model is  a worthwhile endeavor \cite{Rahul}.

\section{Derivation of Eqs.~\eqref{CorrelQu} and \eqref{CorrelAnn}}
  \label{app:details}

In this appendix, we fill  some missing steps in the calculations of the two-time correlation functions. For the deterministic case, we take $\widehat{q}_1$ given by \eqref{eq:q1corrQuen} and substitute it into the second line in (\ref{eq:Q1avQuen}). This gives 
\begin{widetext}
\beq
\langle Q(t_1) \rangle  =  \ii  \int_{-\infty}^\infty  \frac{\widehat{q}_1(k,t_1) - \widehat{q}_1(k,0) }{k}
\frac{dk}{2 \pi} =  \frac{ \rho}{\pi}\int_{-\infty}^\infty dk
\int_0^{t_1}  d\tau \int_\tau^T dt_2   \lambda(t_2){\rm e}^{\omega(k)(2 \tau - t_2 -t_1)}
\eeq
We split the integral over $t_2$ as $\int_\tau^T = \int_\tau^{t_1} + \int_{t_1}^T$,  exchange the order of the integrals over $\tau$  and $t_2$ and evaluate the integrals over $\tau$. This gives
\begin{eqnarray}
  \langle Q(t_1) \rangle  &=&  \frac{ \rho}{\pi}\int_{-\infty}^\infty dk
  \left[
\int_0^{t_1}dt_2 \lambda(t_2) \int_0^{t_2}  d\tau\, {\rm e}^{\omega(k)(2 \tau - t_2 -t_1)}
+ \int_{t_1}^T dt_2 \lambda(t_2)\int_0^{t_1}   d\tau {\rm e}^{\omega(k)(2 \tau - t_2 -t_1)}  \right]\nonumber \\
  &=& \frac{ \rho}{\pi}\int_{-\infty}^\infty dk
  \left[ \int_0^{t_1} dt_2 \, \lambda(t_2)\,
  \frac{{\rm e}^{\omega(k)(t_2 -t_1)} - {\rm e}^{-\omega(k)(t_2 +t_1) }}
       {2 \omega(k)} +  \int_{t_1}^T dt_2 \, \lambda(t_2)\,
       \frac{{\rm e}^{\omega(k)(t_1 -t_2)} - {\rm e}^{-\omega(k)(t_2 +t_1) }}{2 \omega(k)}\right]\nonumber \\
       &=&   \frac{ \rho}{\pi} \int_{0}^T dt_2 \,  \lambda(t_2)
       \int_{-\infty}^\infty dk\,
       \frac{{\rm e}^{-\omega(k)|t_1 -t_2|} - {\rm e}^{-\omega(k)(t_2 +t_1) }}{2 \omega(k)}
\end{eqnarray}
The last integral over $k$ represents a two-time correlation function [cf. Eq.~\eqref{eq:Q1avQuen}]. Subtracting 1 from the first term in the numerator and adding 1 to the second term, we analyze the asymptotic behavior of these two integrals using the same method as in deriving the asymptotic \eqref{VarQquench}. It suffices to identify $2T \to |t_1 -t_2|$ in one integral and $2T \to t_2 +t_1$ in the other. This completes the derivation of Eq.~\eqref{CorrelQu}.

In the annealed case, taking into account $\widehat{q}_1(k,0)$ given by \eqref{init:anneal}, the solution of \eqref{eq:q1source} becomes:
\beq
\widehat{q}_1(k,t) = \frac{  \rho k}{\ii\,  \omega(k) } {\rm e}^{-\omega(k)t}  \int_0^T d\tau\, \lambda(\tau) ({\rm e}^{-\omega(k)\tau} - 1)
     - 2 \ii\,\rho k 
     \int_0^t d\tau \int_\tau^T dt_2\, \lambda(t_2){\rm e}^{\omega(k)(2 \tau - t_2 -t)}
     \label{Ann2terms}
\eeq
The second term on the right-hand side  is the same as in the deterministic case. We only need to evaluate the contribution of the first term. After a bit of algebra we arrive at an integral 
\begin{eqnarray} 
 \rho  \int_0^T dt_2\, \lambda(t_2)  \int_{-\infty}^\infty  \frac{dk}{2 \pi}
  \left(  \frac{1 - {\rm e}^{-\omega(k)t_1} }{  \omega(k) }
    +  \frac{1 - {\rm e}^{-\omega(k)t_2} }{  \omega(k) }
    -  \frac{1 - {\rm e}^{-\omega(k)(t_1 +t_2)} }{  \omega(k) }  \right)
    \label{contribAnn}
\end{eqnarray}
\end{widetext}
When $t_1$ and $t_2$ are large, we use again the same calculation as in deriving the asymptotic \eqref{VarQquench}. We find that the $k$-integral in \eqref{contribAnn} behaves as
\begin{equation*}
W_s
\left(t_1^{\frac{s}{1+s}} + t_2^{\frac{s}{1+s}} - |t_1-t_2|^\frac{s}{1+s} \right)
\end{equation*}
Adding this contribution to the asymptotic of the second term on the right-hand side of \eqref{Ann2terms}, which is just the two-time correlation function in the deterministic case, we arrive at the announced Eq.~\eqref{CorrelAnn}.

\bibliography{references-R}

\end{document}